\documentclass[twocolumn,times,astrosymb]{aastex631}

\usepackage{CJK}
\usepackage{xspace}
\usepackage{amssymb}
\usepackage{xcolor}
\usepackage{multirow}

\graphicspath{{./figures/}}

\shorttitle{Binary Broadening Function}
\shortauthors{Yi}

\begin{document}
\begin{CJK*}{UTF8}{gbsn}

\title{The Binary Broadening Function}

\correspondingauthor{Tuan Yi}
\email{yituan@pku.edu.cn}

\author[0000-0002-5839-6744]{Tuan Yi}
\affiliation{Department of Astronomy, School of Physics, Peking University, Yiheyuan Rd. 5, Haidian District, Beijing, China, 100871}
\affiliation{Kavli Institute of Astronomy and Astrophysics, Peking University, Yiheyuan Rd. 5, Haidian District, Beijing, China, 100871}

\begin{abstract}

We propose an extended formalism for the spectral broadening function (BF) based on the multiplication rule of block matrices.
The formalism, which we named the binary broadening function (BBF), 
can produce decomposed BFs for individual components of a binary star system by using two spectral templates. 
The decomposed BFs can be used to derive precise rotational profiles and radial velocities for individual components.
We test the BBF on simulated spectra and actual observational spectra to show that 
the method is feasible on spectroscopic binaries, even when the spectral lines of two stellar components are heavily blended.
To demonstrate the capability of the method,
we conduct a simulation of `sketching' (imaging) a transiting circumbinary exoplanet using the BBF.
We also discuss issues of implementation such as the variation of BBF with biased templates, 
the pros and cons of BBF, and cases when the method is not applicable.

\end{abstract}

\keywords{
Spectroscopic binary stars (1557), Radial velocity (1332), Close binary stars (254), Stellar spectral lines (1630), Computational methods (1965)
}

\section{Introduction}\label{sec:intro}

The stellar spectral-line broadening is of great importance for measuring the radial velocity (RV) and rotational velocity of stars.
For binary systems with two sets of spectral lines present,  
measuring the parameters of individual components could be a challenging task.
Spectral-line broadening function \citep[BF; ][]{Rucinski1992} is an intuitive method to study such a problem.
The BF is a linear transformation from a so-called sharp-line spectrum (template) to a broad-line spectrum (science/target spectrum).
In essence, the shape-line spectrum (typically the spectrum of a very slow rotating star) 
convolved by the BF provides the best fit to the broad-line spectrum.
As pointed out by \cite{Rucinski1999,Rucinski2002},  
BF has several advantages over the widely used cross-correlation function (CCF), 
as the former preserves the resolution, has a much more clearly defined baseline, 
and is free of the `peaks-pulling' effect as the latter does.

The BF algorithm has been used and explored by several works. 
For example, \cite{Lu1999} and \cite{Rucinski1999b} (followed by a series of fifteen works in total) 
used the BF to measure the RVs of spectroscopic binaries observed by the David Dunlap Observatory (DDO) close binary program.
\cite{Welsh2012} used the BF to measure the RVs of Kepler-34(AB) and Kepler-35(AB), 
two binary systems each containing a pair of solar-like stars with a gas giant planet orbiting around. 
\cite{Cunningham2019} used the BF to extract precise RVs for spectroscopic eclipsing binaries 
from APOGEE spectra to yield more precise binary orbital solutions
(which even led to the discovery of tertiary components in three systems). 
On the other hand, \cite{Yi2022} used the BF to confirm the single-lined spectroscopic binary (SB1) nature of a dynamically measured neutron star candidate,  
and confidently ruled out contamination from a distant third object. 
The preservation of linearity is another strength of BF, making it suitable for reconstructing rotational profiles of stars. 
For instance, \cite{Newton2019} used the BF to measure the projected rotational velocity ($v\sin{i}$) of an exoplanet host star DS Tuc A in a young moving group,
enabling the constraint of the star's inclination and the spin-orbit misalignment. 
\cite{Zheng2022} used the BF to measure the $v\sin{i}$ (for constraining the orbital inclination) for a tidally locked K dwarf, 
of which the companion is a $1.4 - 1.6 M_{\odot}$ neutron star candidate that locates only 127.7 $\pm$ 0.3 pc away from our Sun.
Leaving many more applications unmentioned, 
the BF has shown versatility and flexibility to characterise single, binary, or multi-stellar systems.

In terms of limitations, BF is computationally expensive \citep[much more expensive than CCF;][]{Rucinski1999}.
In addition, BF can only take a single template to represent a double-lined spectrum, 
which could lead to biased parameter estimations when the two components of a binary have a large spectral type deviation.
\cite{Tofflemire2019} explored this issue by examining template selections 
and proposed that the one that yields the smallest RV uncertainty would be the best template,
which typically has stellar parameters close to the dominant component (primary star).
However, the reconstructed broadening profile is inaccurate for the less-dominant star (secondary), 
as it is represented by a biased template.
Therefore when demanding high-precision measurements of parameters for both components of a binary star, 
the original BF method would need certain refinements, as will be presented in this work. 

In the Section~\ref{sec:bbf}, we briefly revisit the formalism of the BF 
and propose an extended formalism, which we named the binary broadening function (BBF for short).
In Section~\ref{sec:test}, we test the BBF on the spectroscopy of a simulated double-lined spectroscopic binary (SB2) 
and the spectra of three actual SB2 systems observed by APOGEE survey.
In Section~\ref{sec:discuss}, we discuss a mock experiment on imaging a circumbinary exoplanet to demonstrate the capability of BBF;
we also discuss issues when using the BBF and the pros and cons of the method.
Conclusions are drawn in Section~\ref{sec:conclude}.

\section{The Binary Broadening Function}\label{sec:bbf}

\subsection{Formalism of the Broadening Function}\label{sec:bbf1}

The linear conversion from a sharp-line template (denoted as $\mathcal{T}$) 
to a broad-line spectrum (denoted as $\mathcal{S}$) is formulated as a system of linear equations:

\begin{equation}\label{eq:bf1}
\hat{\mathrm{D}} \vec{\mathrm{B}} = \vec{\mathrm{S}} \, ,
\end{equation}

where $\hat{\mathrm{D}}$ is an $N \times M$ design matrix that contains 
in each column a velocity-shifted and continuum-normalized  template's flux vector $\vec{\mathrm{T}}$ of length $N$,
$\vec{\mathrm{B}}$ is the BF vector ($M \times 1$), 
and $\vec{\mathrm{S}}$ is the continuum-normalized flux vector ($N \times 1$) of $\mathcal{S}$.  
This concise matrix equation states that the broad-line spectrum $\mathcal{S}$ 
can be expressed as a linear combination of the sharp-line template $\mathcal{T}$ at different velocity shifts,
and the BF is just the coefficients of the combination\footnote{see an intuitive demonstration on: 
\url{https://saphires.readthedocs.io/en/latest/bf_walkthrough.html}}. 
$\vec{\mathrm{B}}$ can also be interpreted as the mean spectral line profile \citep{Tofflemire2019}.
The spectral broadening comes from various sources such as micro-/macro-turbulence, rotation, 
as well as instrumental broadening due to finite resolution of a spectrograph. 
The template $\mathcal{T}$ and the spectrum $\mathcal{S}$ must be re-sampled (re-binned) 
in a uniform velocity space of constant step size, and then rectified/normalized before inserting to the above equation. 
The recipe is by creating a common log-uniform wavelength grid $\log(\lambda_{i}) = \log(\lambda_{0}) + i \delta \log(\lambda) $, 
where $\lambda_{0}$ is the first point, $\delta \log(\lambda)$ is the step size,
and $i = (1, 2, ... N)$ as $N$ being the total length of the spectrum. 
The fluxes are then resampled onto the new wavelength grid, for instance, by using a flux-conservative method \citep[e.g., ][]{Carnall2017}.
After the resampling, the wavelength grid is no longer needed, and one can construct $\hat{\mathrm{D}}$ 
by shifting and plugging $\vec{\mathrm{T}}$ into its columns.

Typically, a piece of spectrum contains a few thousands to tens of thousands of points ($\mathcal{O}(N) = 10^{3} -10^{4}$), 
and the velocity space of interest has hundreds of grid points ($\mathcal{O}(M) = 10^{2}$).
Therefore Equation~(\ref{eq:bf1}) is an overdetermined system of equations.
To solve $\vec{\mathrm{B}}$, \cite{Rucinski1999} proposed an elegant way by using the singular value decomposition (SVD) algorithm.
The SVD decompose $\hat{\mathrm{D}}$  into the product of three matrices, that is, $\hat{\mathrm{D}} = \mathrm{U \Sigma V^{T}}$,
where $\mathrm{U}$ and $\mathrm{V}$ are orthonormal matrices, 
the superscript $\mathrm{T}$ stands for matrix transpose,
and $\mathrm{\Sigma}$ is a diagonal matrix containing singular values in its diagonal.
Therefore, $\vec{\mathrm{B}} =  \hat{\mathrm{D}}^{-1} \vec{\mathrm{S}} =  \mathrm{V \Sigma^{-1} U^{T}} \vec{\mathrm{S}}$, using the fact that the inverse of an orthonormal matrix is its transpose.

\subsection{The Binary Broadening Function}\label{sec:bbf2}

As mentioned in Section~\ref{sec:intro}, 
the standard formalism of BF utilizes one single continuum-normalized template, 
thus may mis-represent systems containing two/multiple components with large spectral type deviations.
Here we introduce a natural formalism to include double templates for the BF,
based on the multiplication rule of block matrix. 
We will use subscript `1'  to stand for the primary and subscript `2' to stand for the secondary throughout the rest of this paper. 
For SB2s,  the problem can be modelled as the following block matrix equation:

\begin{equation}\label{eq:bf2}
\left(
\begin{array}{cc}
\hat{\mathrm{D}}_\mathrm{1} & \hat{\mathrm{D}}_\mathrm{2} 
\end{array}
\right) 
\left(
\begin{array}{c}
\vec{\mathrm{B}}_\mathrm{1} \\ 
\vec{\mathrm{B}}_\mathrm{2}
\end{array}
\right)  
= \hat{\mathrm{D}} \vec{\mathrm{B}} = \vec{\mathrm{S}} \, ,
\end{equation}

where $\hat{\mathrm{D}}_\mathrm{i}$ ($\mathrm{i} =$ 1, 2) are design matrices created from two best-estimated templates 
$\mathcal{T}_{1}$ and $\mathcal{T}_{2}$,
$\vec{\mathrm{B}}_\mathrm{i}$ ($\mathrm{i} =$ 1, 2) are broadening functions for individual components in the SB2.
and  $\vec{\mathrm{S}}$ is our SB2's observational spectrum.

Note that when constructing Equation~(\ref{eq:bf2}), one again has to create a common log-uniform wavelength grid 
onto which the fluxes of both templates and spectrum $\vec{\mathrm{S}}$ will be resampled.
$\hat{\mathrm{D}}_\mathrm{1}$ and $\hat{\mathrm{D}}_\mathrm{2}$ should also have equal number of columns
by applying a same velocity-shifts grid to $\vec{\mathrm{T}}_{1}$ and $\vec{\mathrm{T}}_{2}$. 
The matrix Equation~(\ref{eq:bf2}) can be solved in exactly the same way as Equation~(\ref{eq:bf1}), 
that is, by using the SVD recommended by \cite{Rucinski1999}.
One simply splits the solution $\vec{\mathrm{B}}$ into two halves to obtain the individual BFs for each component.  
We call this extended formalism of the original BF as the binary broadening function (BBF),
although it is easily extensible to triple or even multiple systems with increasing number of block matrices.

\begin{figure*}
	\includegraphics[width=\textwidth]{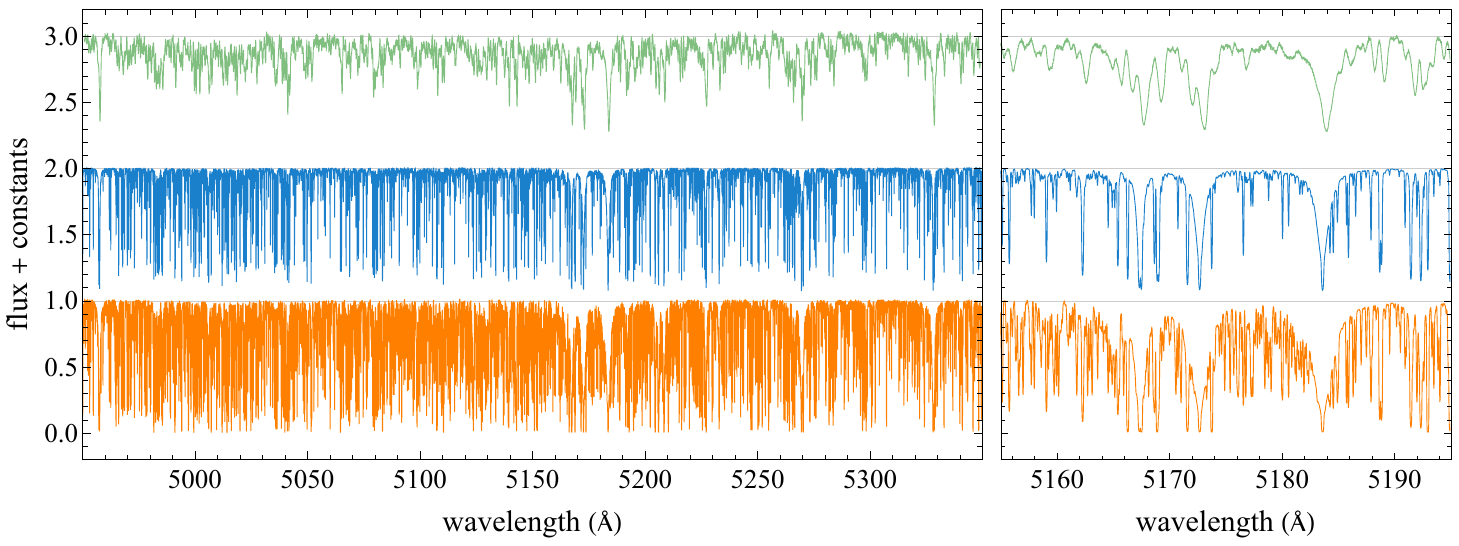}
    \caption{The normalized templates and mock SB2 spectrum used to test our BBF algorithm.
    Left panel: we choose a wavelength window near the vicinity of Mg I b triplets at 5167, 5172, and 5183\AA\ .
    Right panel: zoom-in view in the vicinity of  Mg I b triplets. 
    The blue and the orange spectra are the sharp-line templates used for the primary and the secondary, respectively.
    The green spectrum is a simulated broad-line spectrum for a mock SB2 (see texts for the details).
    The spectra are vertically shifted with a constant for clarity.}
    \label{fig:mock12}
\end{figure*}

\begin{figure}
\centering
	\includegraphics[width=1.0\columnwidth]{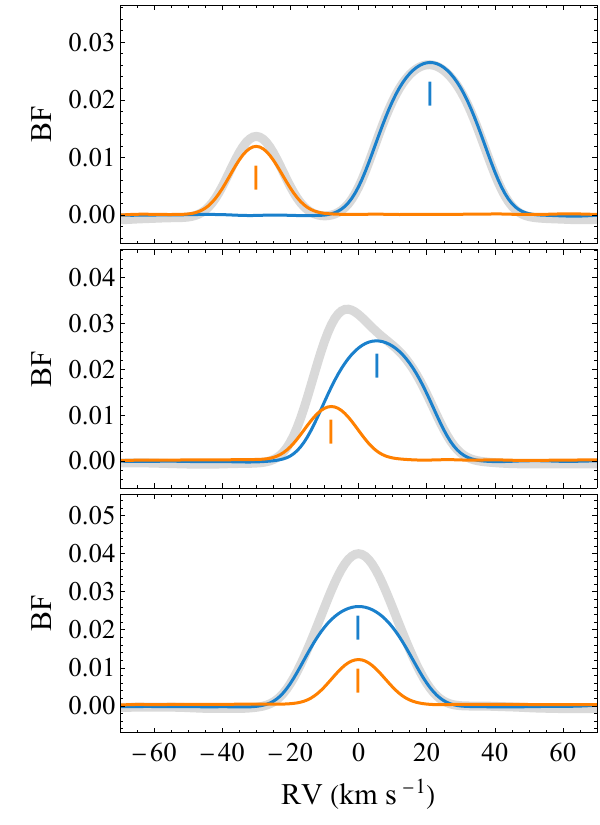}
    \caption{The BBF solutions for our SB2 with three different mock spectra taking at at different orbital phases.
    The blue curve and orange curve are the individual BF for the primary ($\mathrm{BF}_{1}$) and the secondary ($\mathrm{BF}_{2}$), respectively.
    Vertical color bars indicate the RVs for the two components.
    As for comparison, the shallow grey line stands for the solution of the original BF 
    using a single template (in this case $\mathcal{T}_{1}$) to formulate the problem. 
    Three panels show solutions when the two components has clearly separated BF peaks (upper panel),
    partially blended peaks (middle panel), and totally blended peaks (lower panel).  
    In all cases, we obtained well decomposed BFs for individual components.
    }
    \label{fig:bf123}
\end{figure}

\begin{figure}
\centering
	\includegraphics[width=1.0\columnwidth]{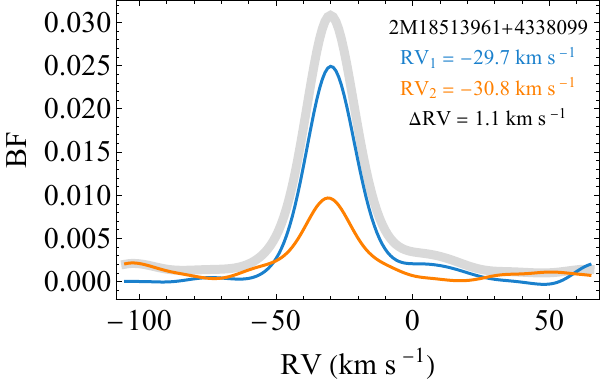} \\
	\includegraphics[width=1.0\columnwidth]{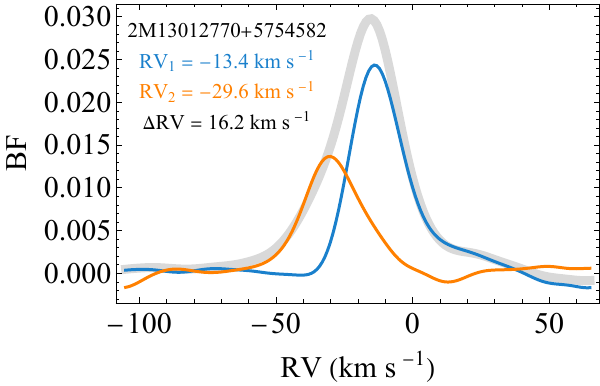} \\
	\includegraphics[width=1.0\columnwidth]{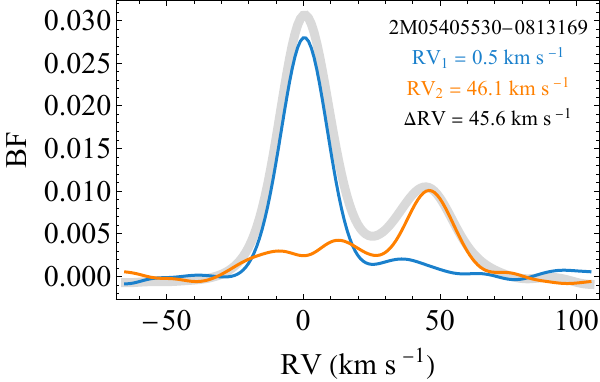} 
    \caption{The BBFs for the APOGEE spectra of 
    2M18513961+4338099 (upper panel), 2M13012770+5754582 (middle panel), and 2M05405530-0813169 (lower panel).
    The blue curve and orange curve respectively stand for the individual BFs for the primary and secondary components in the system. 
    The single-template BF is shown by the shallow grey curve for comparison.
    RV offsets derived from the BBF are consistent with the results given by \cite{El-Badry2018}. }
    \label{fig:apogee}
\end{figure}

\section{Testing the BBF}\label{sec:test}

\subsection{A Mock SB2 System}\label{sec:sb2}

To test the BBF,  we create mock SB2 spectroscopy using the BOSZ stellar atmospheric models
\footnote{\url{https://archive.stsci.edu/prepds/bosz/}}.
We adopt models with spectral resolution  $R =$ 300\,000, 
default microturbulent velocity $= 2.0~\mathrm{km~s^{-1}}$ and zero rotational broadening velocity.  
For simplicity, only models with zero alpha and carbon abundance are used.
Therefore, the parameter space is spanned only by 
the effective temperature $T_\mathrm{eff}$,  surface gravity $\log{g}$, and the metallicity $\mathrm{[Fe/H]}$.

We simulate a mock SB2 consisting of a solar-like primary characterized with parameters:
$T_\mathrm{eff~1} = 5800~\mathrm{K}$, $\log{g}_{1} =  4.4~\mathrm{dex}$,  $\mathrm{[Fe/H]}_{1} = 0.0~\mathrm{dex}$, 
and a cooler subgiant secondary characterized with parameters:
$T_\mathrm{eff~2} = 4500~\mathrm{K}$, $\log{g}_{2} = 2.8~\mathrm{dex}$,  $\mathrm{[Fe/H]}_{2} = 0.0~\mathrm{dex}$.
The corresponding spectra for the primary (denoted as $\mathcal{T}_{1}$) and the secondary  (denoted as $\mathcal{T}_{2}$) 
are linearly interpolated from the BOSZ synthetic spectral model grid described in the last paragraph.
$\mathcal{T}_{1}$ and $\mathcal{T}_{2}$ are then resampled and normalized to construct the design matrices (Section~\ref{sec:bbf2}).
In addition, we assume that the projected equatorial rotational velocities for the primary and the secondary are 
$v\sin{i}_{1} = 20~\mathrm{km~s^{-1}}$ and $v\sin{i}_{2} = 10~\mathrm{km~s^{-1}}$, respectively. 
\footnote{Note that the assumed stellar parameters may not be physically plausible, but would be handy for the purpose of testing.}
Since SB2s are radial velocity variables, 
we will test a set of velocity pairs $\mathrm{RV_{1}}$ and $\mathrm{RV_{2}}$ to mimic spectra observed at different orbital phases.

To generate a combined SB2, each template is first shifted to its corresponding $\mathrm{RV}$, 
broadened with a rotational profile \citep{Gray2005} for a given $v\sin{i}$ 
(assuming zero limb-darkening for convenience), and flux co-added. 
We then broaden the combined spectrum with a gaussian kernel corresponding to a spectral resolution $R$ = 30\,000 to mimic the instrumental broadening of a virtual spectrograph.
Last, we normalize the spectrum and multiply the fluxes by random fluctuations following a gaussian distribution $\mathcal{N}(1,0.01)$ 
(gaussian centered at one with a standard deviation of 0.01) 
to mimic a high-quality observational spectrum with a signal-to-noise-ratio (SNR) $\sim$ 100.
Figure~\ref{fig:mock12} shows the normalized templates for individual components and the simulated SB2 spectrum. 

We tested the BBF solution for three mock spectra presumably taking at different orbital phases of the SB2.
For instance, at quadrature phases where two stars are side-by-side with respect to an observer, 
the BF has well-separated peaks. 
On the contrary, near conjunction phases when the two stars align with the line-of-sight 
(assuming no eclipsing for convenience), 
the BF manifests partially or totally blended peaks.
The solutions for these cases are shown in Figure~\ref{fig:bf123}.  
For clear RV separated case (upper panel in Figure~\ref{fig:bf123}), 
the BBF solution provides clear decomposition of the individual BFs. 
In particular the secondary has now represented by its own well-matched template,
therefore its component $\mathrm{BF_{2}}$ makes more physical sense for either broadening or the relative line width.
For partially or even totally blended cases (lower and bottom panels in Figure~\ref{fig:bf123}), 
we again obtained well-separated BFs for each component.
This is a useful property for the BBF when one does not know which broadening mechanism dominates the spectrum,
for instance, rotational broadening or instrumental  broadening 
(typically, the former is a Gray rotational profile \citep{Gray2005} and the latter could be a gaussian profile). 
In this case, one cannot simply fit the single-template BF with a specific model to decompose the BF,
therefore the BBF helps to reconstruct the underlying profile.

\begin{table*}
\renewcommand{\arraystretch}{1.05}	
\centering
  \caption{Comparisons of the RVs and RV offsets for three APOGEE SB2s}
  \centering
  \footnotesize
	\begin{tabular}{@{}lllll@{}} 
	\hline
	\textbf{Target ID}  &  \textbf{Parameter}     & \textbf{BF} & \textbf{BBF} & \textbf{\cite{El-Badry2018}}  \\
	(1)               	    & (2) & (3) & (4) & (5) \\
	\hline
	\multirow{3}{*}{2M18513961+4338099}
		 & 	$\mathrm{RV}_{1}$ ($\mathrm{km~s^{-1}}$)     & $-$29.9	             & $-$29.7	 & not given	  \\ 
		 & 	$\mathrm{RV}_{2}$ ($\mathrm{km~s^{-1}}$)     & not measured.    & $-$30.8	 & not given 	  \\ 
		 & 	$\Delta \mathrm{RV}$ ($\mathrm{km~s^{-1}}$) & not measured     & 1.1	         & 1.7	           \\
	\hline
	\multirow{3}{*}{2M13012770+5754582}
		 & 	$\mathrm{RV}_{1}$ ($\mathrm{km~s^{-1}}$)      &  $-$15.7             & $-$13.4	 & not given	 \\ 
		 & 	$\mathrm{RV}_{2}$ ($\mathrm{km~s^{-1}}$)      &  not measured   & $-$29.6	 & not given 	  \\ 
		 & 	$\Delta \mathrm{RV}$ ($\mathrm{km~s^{-1}}$)  &  not measured   & 16.2	         & 16.1	          \\
	\hline
	\multirow{3}{*}{2M05405530-0813169}
		 & 	$\mathrm{RV}_{1}$ ($\mathrm{km~s^{-1}}$)     &  0.7	             & 0.5	         & not given	 \\ 
		 & 	$\mathrm{RV}_{2}$ ($\mathrm{km~s^{-1}}$)     &  43.6                   & 46.1	         & not given 	 \\ 
		 & 	$\Delta \mathrm{RV}$ ($\mathrm{km~s^{-1}}$) &  42.9                   & 45.6	         & 44.9	         \\
	\hline
  \end{tabular}\label{tab:drv}
  
  \vspace{1ex}
  {\raggedright 
  Notes: the RVs and RV offsets for three APOGEE SB2s are
  measured by the single-template BF method (column 3), 
  by the BBF method in this work (column 4), 
  and adopted from \cite{El-Badry2018}'s work (column 5). 
  Note that the $\mathrm{RV_{2}}$ and RV offset for 2M13012770+5754582 and 2M18513961+4338099 are not measured for the BF method,
  since the spectral lines of the secondary star are heavily blended with that of the primary, 
  leaving only a measurable primary peak on the BF (Figure~\ref{fig:apogee}). 
 \par} 
\end{table*}

\subsection{Three Double-lined Spectroscopic Binaries from APOGEE}\label{apogge3}

In this section we apply the BBF on three SB2s to measure the RVs and RV offsets for the components;
they are: 2M18513961+4338099, 2M13012770+5754582, and 2M05405530-0813169.
These are examples presented by the work of \cite{El-Badry2018} (Section 3.1) to demonstrate 
a flexible data-driven method for identifying and fitting the spectra of binary-/multiple-star systems observed by the APOGEE survey \citep{Majewski2017}.
Both three systems are main-sequence binaries containing a primary star with 
$T_\mathrm{eff} \approx 5400$K, $\log{g} \approx 4.5$ dex, $\mathrm{[Fe/H] \approx 0.0}$ dex,
and a mass ratio $q = M_{2}/M_{1} \approx 0.7$ 
(where $M_{1}$ and $M_{2}$ denote the mass for the primary and the secondary, respectively).
The RV offsets for the primary and the secondary star in the spectra of 
2M18513961+4338099, 2M13012770+5754582, and 2M05405530-0813169 respectively are:
$\Delta{\mathrm{RV}}= 1.7, 16.1$, and  $44.9  \mathrm{km~s^{-1}}$  \citep[][Figure 2]{El-Badry2018}.
As pointed out by \cite{El-Badry2018}, the spectral resolution of APOGEE spectra  is $R \sim 22500$, 
which is equivalent to an RV difference of $\delta v \sim c/ R \sim 13.5 \mathrm{km~s^{-1}}$.
Components with a small velocity offset (e.g., $\delta v \lesssim 30 \mathrm{km~s^{-1}}$ ) can not be reliably modelled and fitted with traditional methods \citep{Fernandez2017}. 
Therefore the three systems are good examples to test and demonstrate the feasibility of the BBF algorithm.

The APOGEE detector has three wavelength regions: 
1.514 -- 1.581 $\mu$m, 1.585 -- 1.644 $\mu$m, and 1.647 --1.696 $\mu$m. 
In this work, we adopt all three regions to calculate the BBF.
APOGEE spectra contain a substantial amount of artifact due to poorly corrected telluric lines, 
which are visually identified and removed.
Then the spectra are rectified by fitting and dividing out a fourth-order polynomial to the highest flux points of a smoothed version of the spectra.
For the primary's template, we adopt a BOZS template with 
$T_\mathrm{eff} = 5400$ K, $\log{g} = 4.5$ dex, and $\mathrm{[Fe/H] = 0.0}$ dex;
for the secondary' s template, we fix the surface gravity $\log{g} = 4.7$ dex and metallicity $\mathrm{[Fe/H] = 0.0}$ dex,
while setting the effective temperatures to be  $T_\mathrm{eff} = $ 3800, 4050, and 4200 K 
for 2M18513961+4338099, 2M13012770+5754582, and 2M05405530-0813169, respectively
(the choice of  $T_\mathrm{eff} $ is discussed in Section~\ref{sec:temp}).
The BF and the BBF for three systems are calculated and shown in Figure~\ref{fig:apogee}.
As one can see, the BF for individual components are decomposed successfully for all three systems.
We estimate the RVs for the primary and the secondary
by fitting a theoretical rotational profile \citep{Gray2005} to the top of the BFs,
and obtained RV offsets for the three systems: $\Delta{\mathrm{RV}}= 1.1, 16.2$, and $45.6  \mathrm{km~s^{-1}}$, respectively.
The RV offsets are consistent with the results derived by \cite{El-Badry2018}.
Table~\ref{tab:drv} lists the RVs and RV offsets for the three APOGEE SB2s,
measured by the single-template BF, by the BBF, and adopted from \cite{El-Badry2018}'s work. 
Note that the $\mathrm{RV_{2}}$ and RV offset for 2M13012770+5754582 and 2M18513961+4338099 are not measured for the BF method,
since the spectral lines of the secondary star are heavily blended with that of the primary, 
leaving only a measurable primary peak on the BF (Figure~\ref{fig:apogee}).

\begin{figure*}
\centering
	\includegraphics[width=0.24\textwidth]{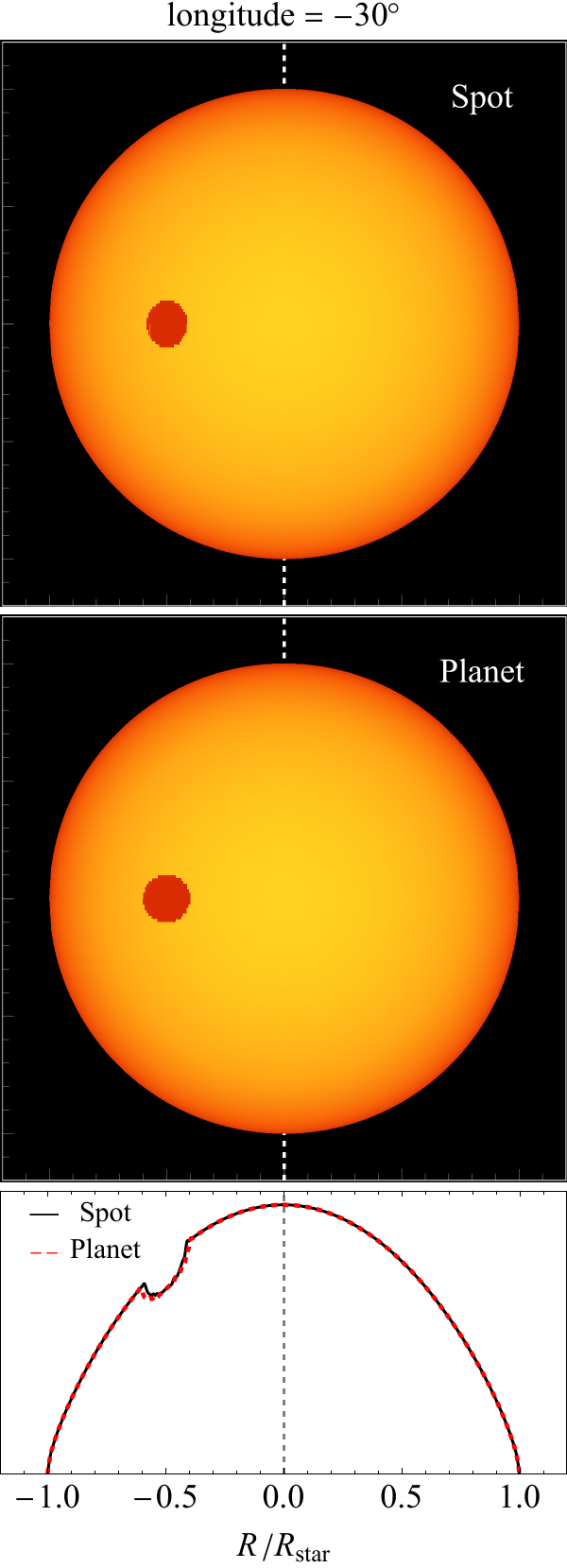}
	\includegraphics[width=0.24\textwidth]{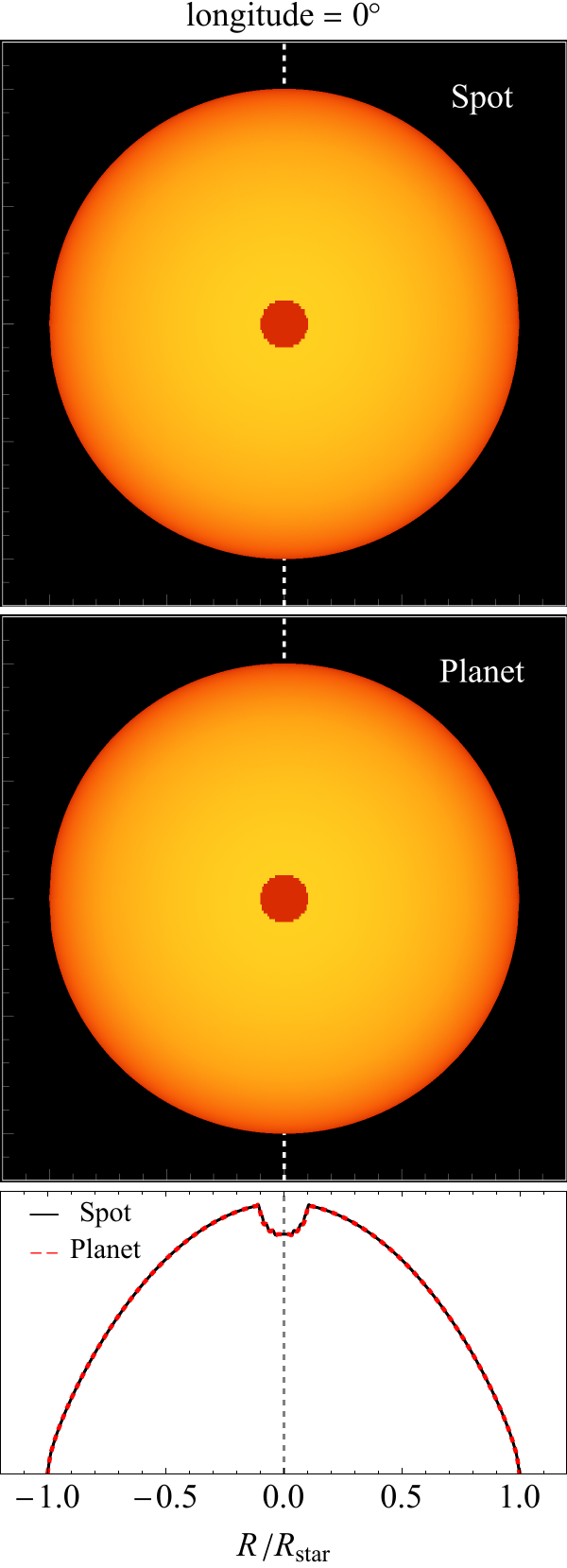}	
	\includegraphics[width=0.24\textwidth]{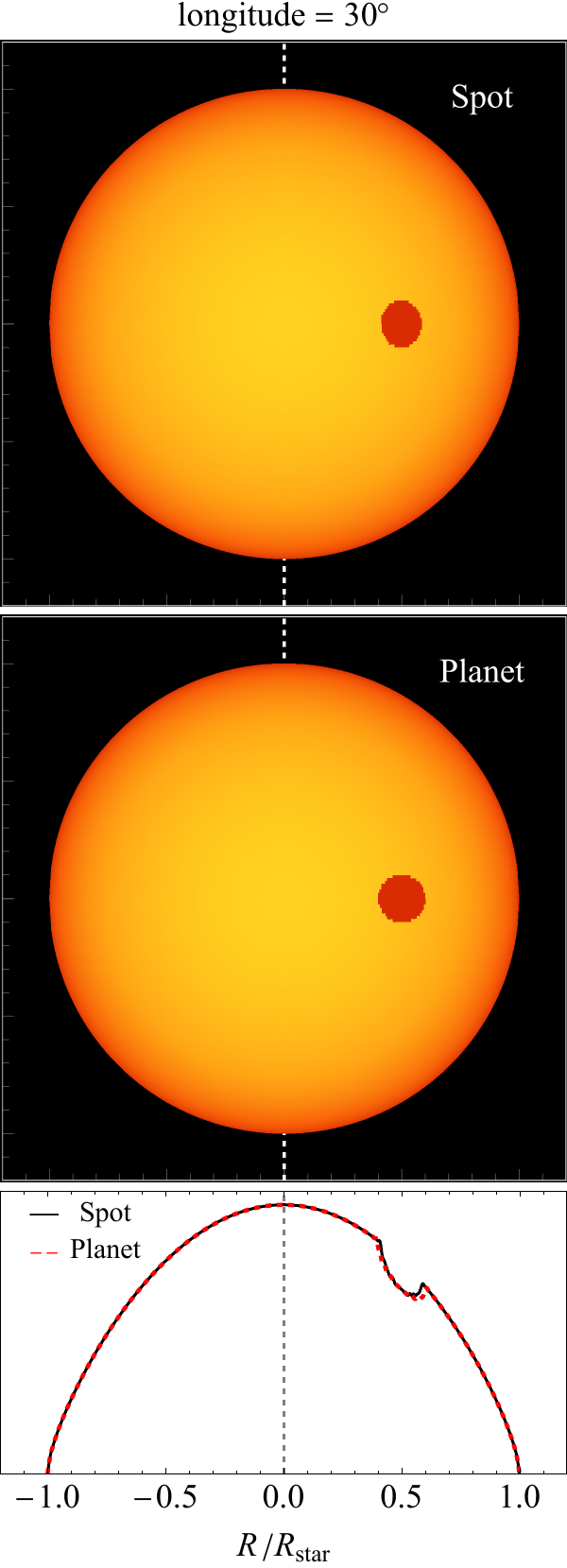}	
	\includegraphics[width=0.24\textwidth]{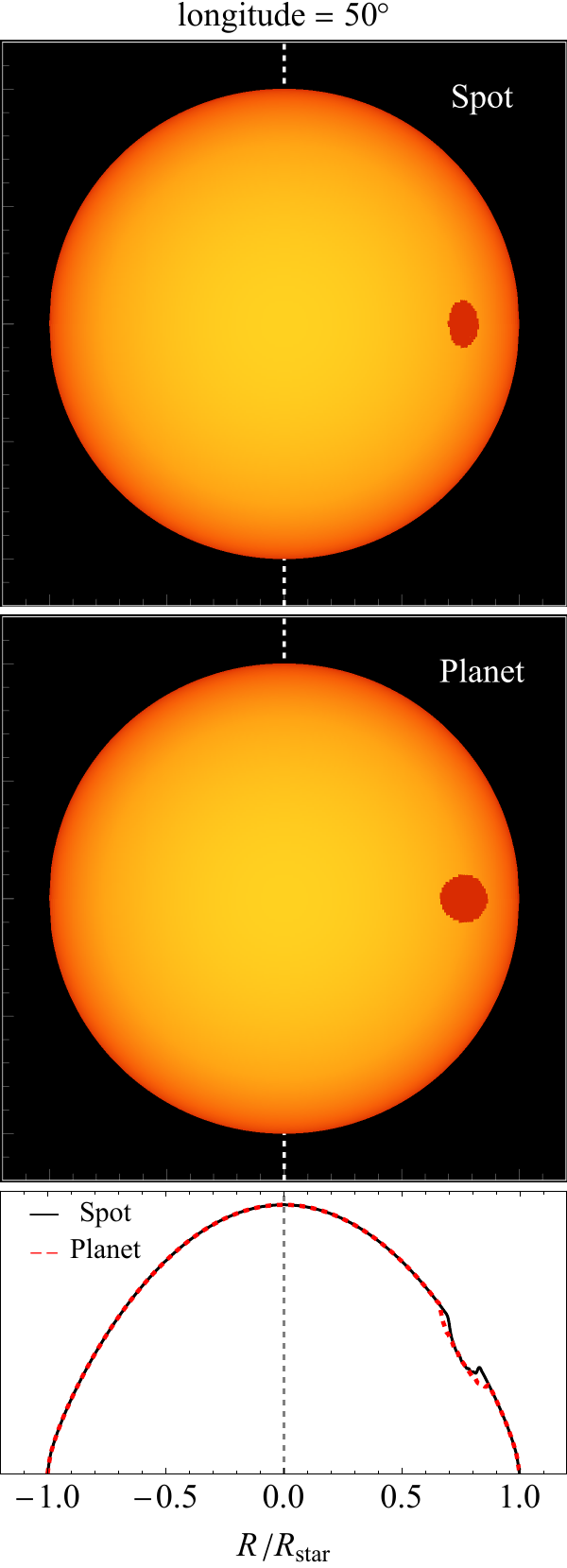}	
    \caption{The physical model for a stellar spot migrating across the surface of a red dwarf
    (first row) and for a planet transiting the red dwarf (second row).
    The star is rotating from left to right with a rotation axis lying within the screen in the vertical direction (white dashed line).
    The color variation indicates the stellar surface brightness variation given the limb-darkening law: yellow for brighter regions and red for darker regions.
    The bottom row shows the BF of the red dwarf, for the spot case (black line) and the exoplanet case (red dashed line). 
    Either spot or planet induces a notch (dip) on the BF which migrates along the BF from left to right.
    The difference can be told when the spot or the projection of the planet lies off-center of the surface:
    moving towards the limb of the star the projection becomes an ellipse for the spot and would still be a circle for the planet. 
    Thus by carefully examining/modelling the shape of the notch, one might be able to identify the planet signal apart from a spot. 
    }
    \label{fig:bfcompare}
\end{figure*}

\begin{figure*}
	\includegraphics[width=\textwidth]{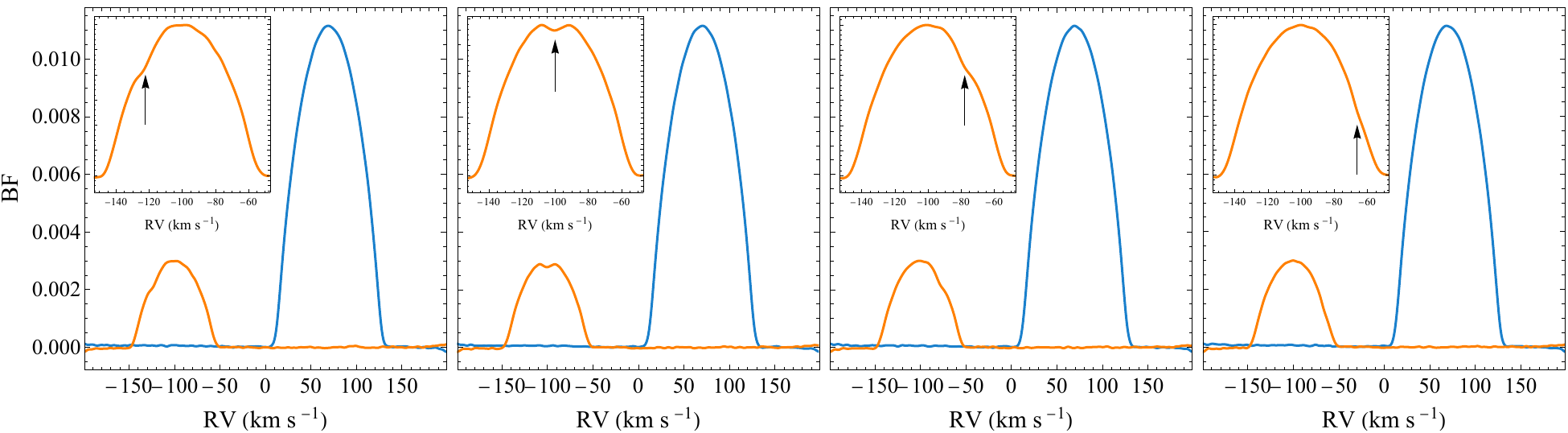} 
    \caption{The BBFs for the spectra of a mock binary taken at four different epochs during the circumbinary exoplanet is transiting the secondary. 
    The relative position of the planet to the secondary star for the four panels corresponds to the four columns in Figure~\ref{fig:bfcompare}.
    In each panel, the blue curve is the decomposed BF for the primary star and the orange curve is for the secondary.
    The inset panel shows a zoom-in view of the secondary's BF. The Notch induced by the transiting planet is indicated by the black arrow. 
    The results show that the BBF unambiguously identifies that the transit signal is associated with a planet passing in front of the secondary star.}
    \label{fig:bbfcbp}
\end{figure*}

\section{Discussion}\label{sec:discuss}

In these section, we discuss potential applications and technical aspects for using the BBF.

\subsection{`Sketching' Circumbinary Exoplanets/Substellar Companions Using BBF}
Since the first evidence of exoplanets orbiting binary systems -- the circumbinary exoplanets system Kepler-16 -- 
has been discovered in 2010 \citep{Doyle2011},
today over a dozen of binary stars with transiting exoplanets have been found and confirmed.
Modelling and inferring the rotation and surface features of these binary systems is a demanding problem.
For instance, stellar activities such as the migration of spots across the stellar surface should be carefully distinguished from an exoplanet transit signal.
In this section, we present simulations to study the BF of stellar spectra presumably taken 
during a transiting event by an exoplanet/substellar companion (e.g., brown dwarf) .
The simulation compares the shapes of the BF either caused by a transiting exoplanet or a rotating spot. 
Then we simulate the spectra of a binary system with transiting exoplanets and 
discuss the possibility to `sketching' (imaging/charaterizing) circumbinary exoplanets using the BBF.

Below are some assumptions for the simulations.
(A) Assumptions for the stars:
(1) the binary system is comprised of a red dwarf and a yellow dwarf (of which the parameters will be given shortly); 
(2) we assume that both stars are spherical and neglect the Roche geometry for simplicity;
(3) a linear limb-darkening law \citep[stellar disk's center-to-limb brightness variation; ][]{Claret2011} 
is adopted with a linear limb-darkening coefficient = 0.7 for both stars;
(4) the orbit of the binary is edge-on, meaning that the orbital inclination is 90 degree;
(5) there is no mutual eclipsing between two stars, that is, the spectrum are taken when the two stars are far apart along the line-of-sight. 
(B) Assumptions for the planet/substellar companion (hereafter the planet): 
(1) the planet is spherical and it is transiting the red dwarf with a radius $R_\mathrm{star} =  0.7 R_{\odot} $;
the planet's radius $R_\mathrm{p} = 0.1 R_\mathrm{star}$; 
(2) the orbit of the planet, the equator of the stars, and the orbit of the binary are all confined perfectly within a plane;
that is, there is NO spin/rotation-to-orbit misalignments. 
(3) the planet has no emission and reflects no light from the star.
(C) Assumptions for the dark/cold stellar spot:
(1) the stellar spot has a circular shape and emits no (negligible) light compared to its ambient surface, that is, it is a `perfectly' dark spot;
(2) the radius of the stellar spot $R_\mathrm{s} = 0.1 R_\mathrm{star}$, comparable to that of the planet for the sake of comparison;
(3) the spot sits at the equator of the red dwarf and migrates across the stellar surface following the rotation of the star, 
with no variation of either size, shape, or brightness.

Figure~\ref{fig:bfcompare} shows the physical model for our simulation. 
The first row are four epochs of a stellar spot that is migrating across the surface of the red dwarf
and the second row are four epochs of a planet transiting the red dwarf.
The star is rotating from left to right with a rotation axis lying within the screen in the vertical direction (white dashed line).
The color variation indicates the stellar surface brightness variation
given the limb-darkening law: yellow for brighter regions and red for darker regions.
The bottom row shows the BF of the red dwarf, for the spot case (black line) and the exoplanet case (red dashed line). 
Note that the horizontal positions of the stellar disk and its broadening profile has a one-to-one correspondence \citep[][page 469]{Gray2005}.
Here the BF is numerically computed by slicing the stellar disk into strips
parallel to the axis of rotation and then by integrating the flux of each strip
(assuming that the surface within each strip have a same RV value when the strip width is sufficiently thin). 
It is clear that either spot or planet induces a notch (dip) on the BF which migrates along the BF from left to right.
The difference can be told when the spot or the projection of the planet lies off-center of the stellar disk:
moving towards the limb of the star the projection becomes an ellipse for the spot and would still be a circle for the planet. 
Hence by carefully examining/modelling the shape of the notch, one might be able to identify the planet signal apart from a spot. 
Our model is an oversimplified toy one for the purpose of demonstration.
For instance, the shape of the spot are almost certainly not a perfect circle, which might also vary across the surface when migrating.
Thus a notch induced by a spot could show much more variations.  
For a transiting planet however, the shape of the notch follows a well defined pattern at different epochs, 
thus multi-epoch spectroscopy would be needed for imaging an exoplanet transiting event.

Now we put the planet + red dwarf described above into a mock binary system and 
use the BBF to recover the notch induced by the transiting planet. 
The red dwarf is the secondary star in the system and has the following parameters:  
$T_\mathrm{eff} = 4200$ K, $\log{g} = 4.5$ dex, $\mathrm{[Fe/H] = 0.0}$ dex, $M = 0.7 M_{\odot}$, and $R = 0.7 R_{\odot}$.
The primary is a yellow dwarf with the following parameters:
$T_\mathrm{eff} = 5400$ K, $\log{g} = 4.6$ dex, $\mathrm{[Fe/H] = 0.0}$ dex, $M = 0.9 M_{\odot}$, and $R = 0.9 R_{\odot}$.
We assume that the binary orbital period $P_\mathrm{orb} = 0.8$ day and the rotation of two stars are synchronized.
The rotational broadening is calculated using $v_\mathrm{rot} = 2 \pi R_\mathrm{star} / P_\mathrm{orb}$,
that is,  $v_\mathrm{rot} \approx 57$ and $44 \mathrm{km~s^{-1}}$ for the primary and the secondary, respectively. 
The mock spectra are generated using exactly the same method described in Section~\ref{sec:sb2},
except that we now simulate a higher SNR of 500 at higher spectral resolution of 50000 
\footnote{$R$ = 50000 corresponds to a resolution of $\sim 6 \mathrm{km~s^{-1}}$ in velocity space},
and the secondary is broadened by a numerical rotational kernel with a transit.
Four different epochs during the transit is simulated as described in the previous paragraph.
Figure~\ref{fig:bbfcbp} shows the result of decomposed BBF for the binary.     
In each panel, the blue curve is the decomposed BF for the primary star and the orange curve is for the secondary.
The inset panel shows a zoom-in view of the secondary's BF. 
Notches induced by the transiting planet is indicated by the black arrow. 
The results show that the BBF unambiguously identifies that the transit signal comes from a planet transiting the secondary star.
As demonstrated, the BBF could be a useful tool to obtain Doppler imaging of circumbinary planets, using high-resolution multi-epoch spectroscopy.

Note that the application of the BBF is not necessarily confined to gravitationally bounded binary stars.
Another potential application might be using the BBF to decompose the spectrum of microlensing events
caused by stars happening to align with each other along the line-of-sight \citep{dong2019}.
In most cases, the spectrum of a microlensing event is not decomposable, 
because the background (source) star is typically much brighter than the foreground (lens) star.
However, there are a few cases when both the source and lens can be resolved via spectroscopy. 
One such example is Gaia22dkv \citep{Wu2023}, a microlensing event alerted by the Gaia team \citep{Gaia2016},
where the source is a red giant and the lens is a main-sequence turn-off star.
\cite{Wu2023} have discovered an exoplanet hosted by the foreground star.
Gaia22dkv is the most promising microlensing planetary event to be characterized by RV follow-up observations.
In particular, to decompose the source and the lens's spectra, to make accurate RV measurements for the lens star,
and to dynamically constrain the planet's mass.
We plan to propose follow-up spectroscopy and implement the BBF method for Gaia22dkv in a future work.

\begin{figure*}
	\includegraphics[width=\textwidth]{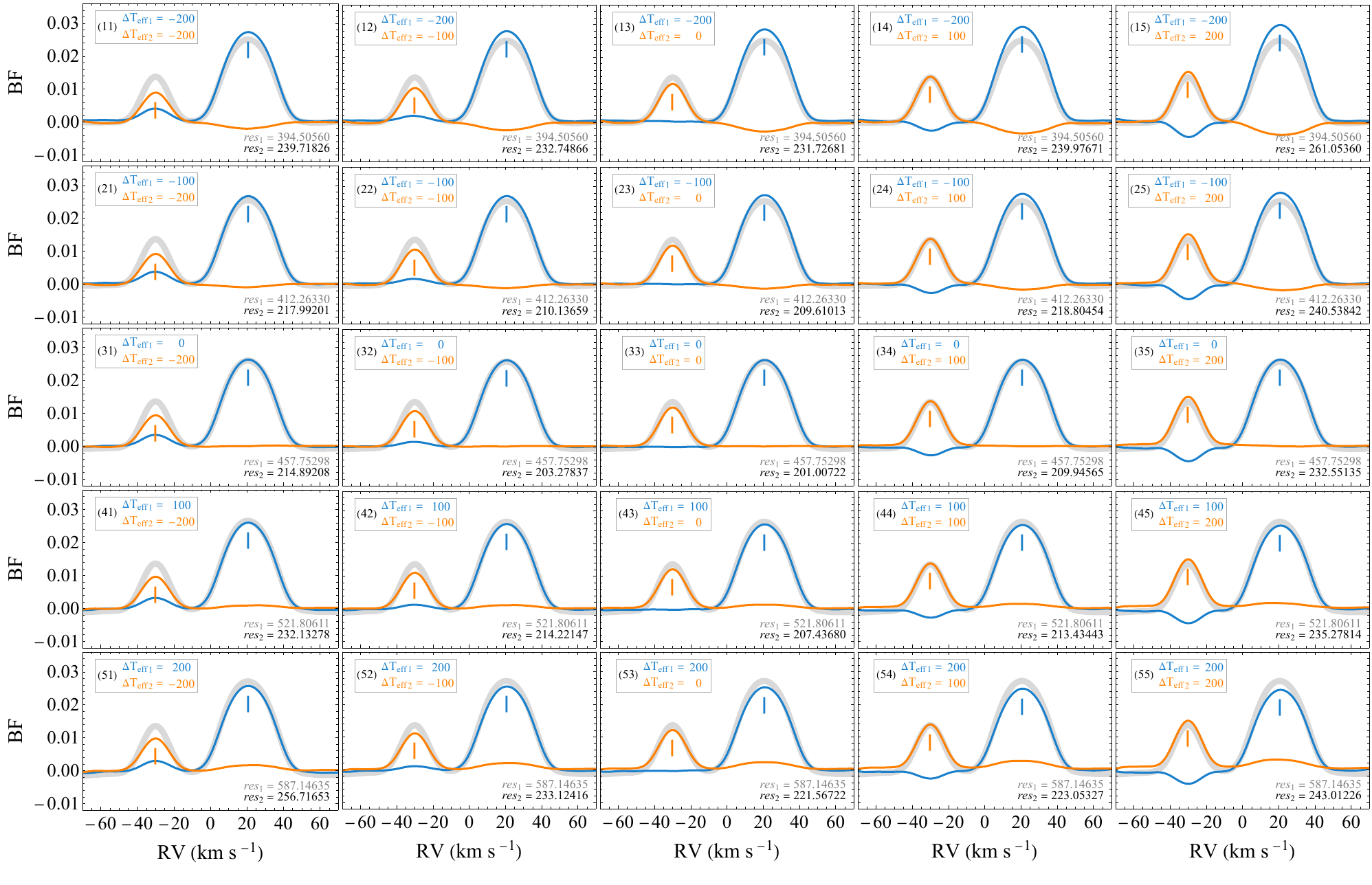}
    \caption{BBF and BF with varying $T_\mathrm{eff}$s for mock SB2's primary and secondary's templates. 
    Lines and colours hold same meanings as that of Figure~\ref{fig:bf123}.
    The deviation of $T_\mathrm{eff}$s from the best-fit values are indicated by $\Delta T_\mathrm{eff}$ in panels' legends:
    minus/plus sign indicates an under/overestimate of the $T_\mathrm{eff}$. 
    By observing the pattern through the third row of panels, 
    one can find that under/overestimating the $T_\mathrm{eff~2}$ induces a bump/pit on $\mathrm{BF_{1}}$ right below the peak of $\mathrm{BF_{2}}$;
    and by observing the pattern through the third column of panels,  
    one can find that under/overestimating the $T_\mathrm{eff~1}$ induces a bump/pit on $\mathrm{BF_{2}}$ right below the peak of $\mathrm{BF_{1}}$.
    }
    \label{fig:teff}
\end{figure*}

\begin{figure*}
	\includegraphics[width=\textwidth]{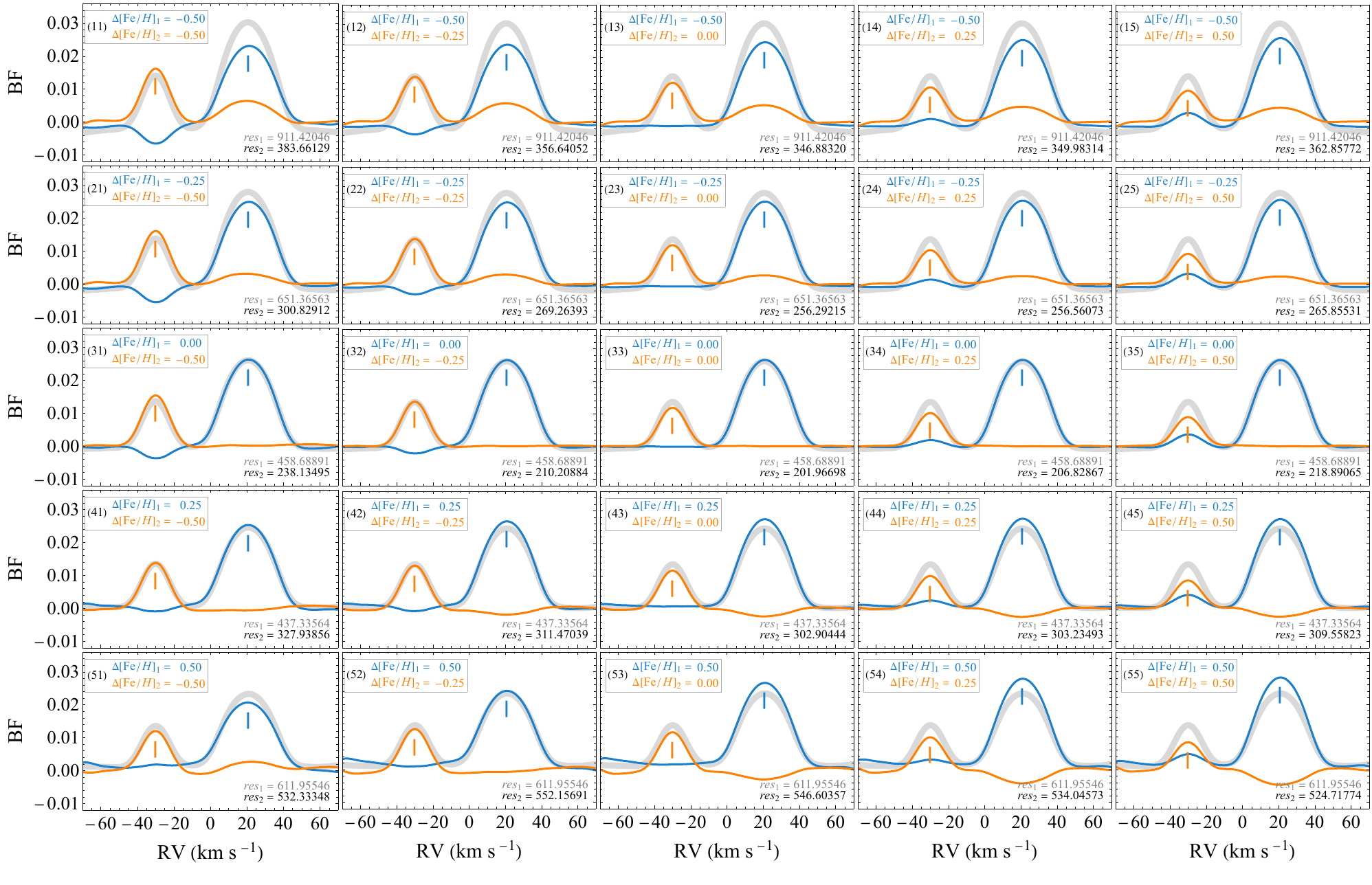}
    \caption{Same as Figure~\ref{fig:teff} but for varying $\mathrm{[Fe/H]}$s.}
    \label{fig:meta}
\end{figure*}

\begin{figure*}
	\includegraphics[width=\textwidth]{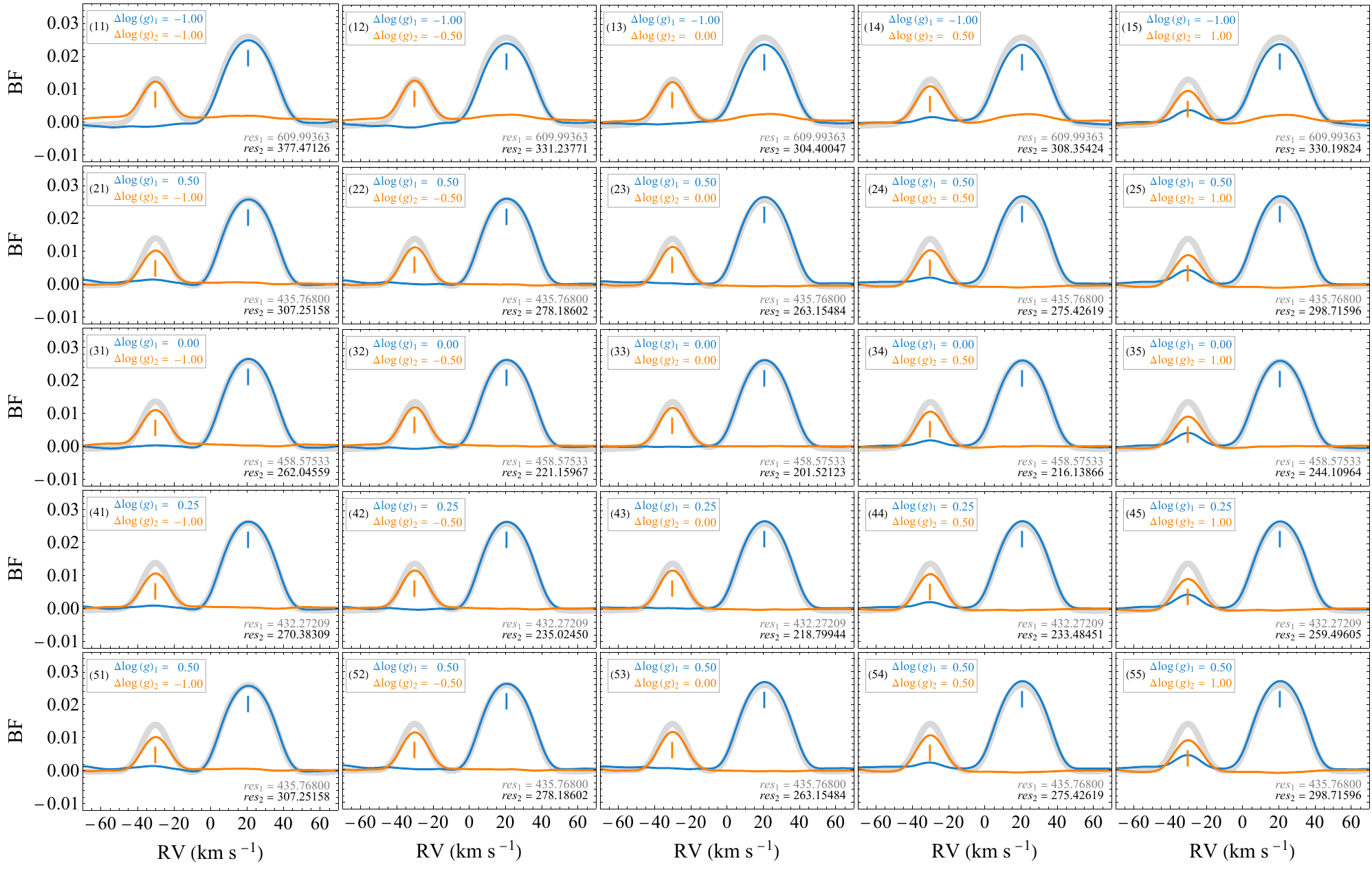}
    \caption{Same as Figure~\ref{fig:teff} but for varying $\log{g}$s.}
    \label{fig:logg}
\end{figure*}

\begin{figure*}
	\includegraphics[width=\textwidth]{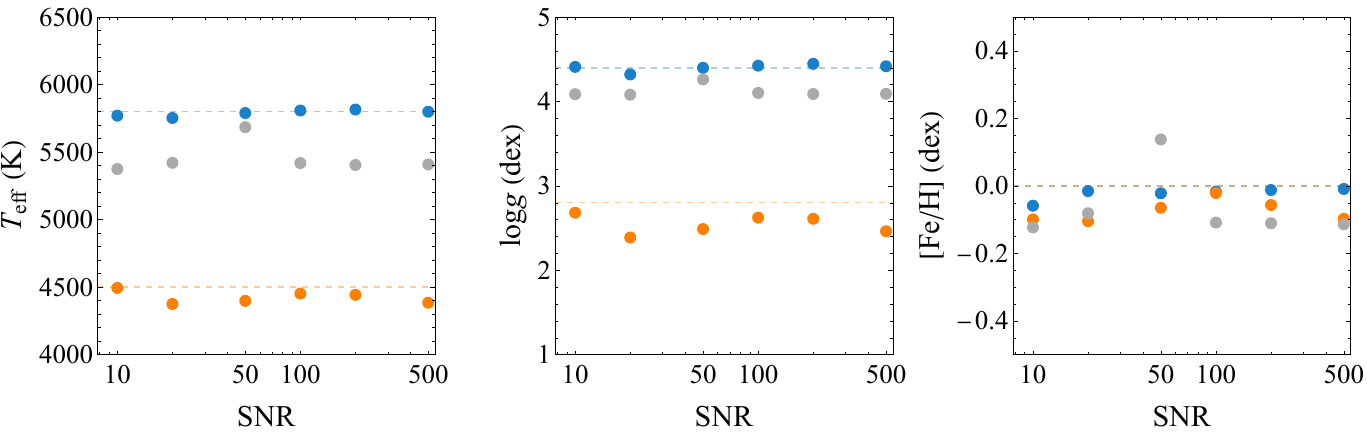}
    \caption{Optimization on templates' parameters using the BBF wrapped up with the Nelder-Mead algorithm.
    The blue and orange dashed lines represent the true parameters for the primary and secondary respectively,
    and the blue and orange points represent the optimized parameters for the primary and the secondary respectively. 
    The results converge to values in agreement with the correct parameters used to simulate the SB2.
    In comparison, we run a parallel test using the single template BF wrapped up with the Nelder-Mead algorithm.
    The best single-template's parameters (grey points) tend to run close to that of the primary,
    The result best-fit single template does not represent secondary nor primary, 
    but instead a flux-weighted template from the two components.}
    \label{fig:optim}
\end{figure*}

\begin{figure*}
	\includegraphics[width=\textwidth]{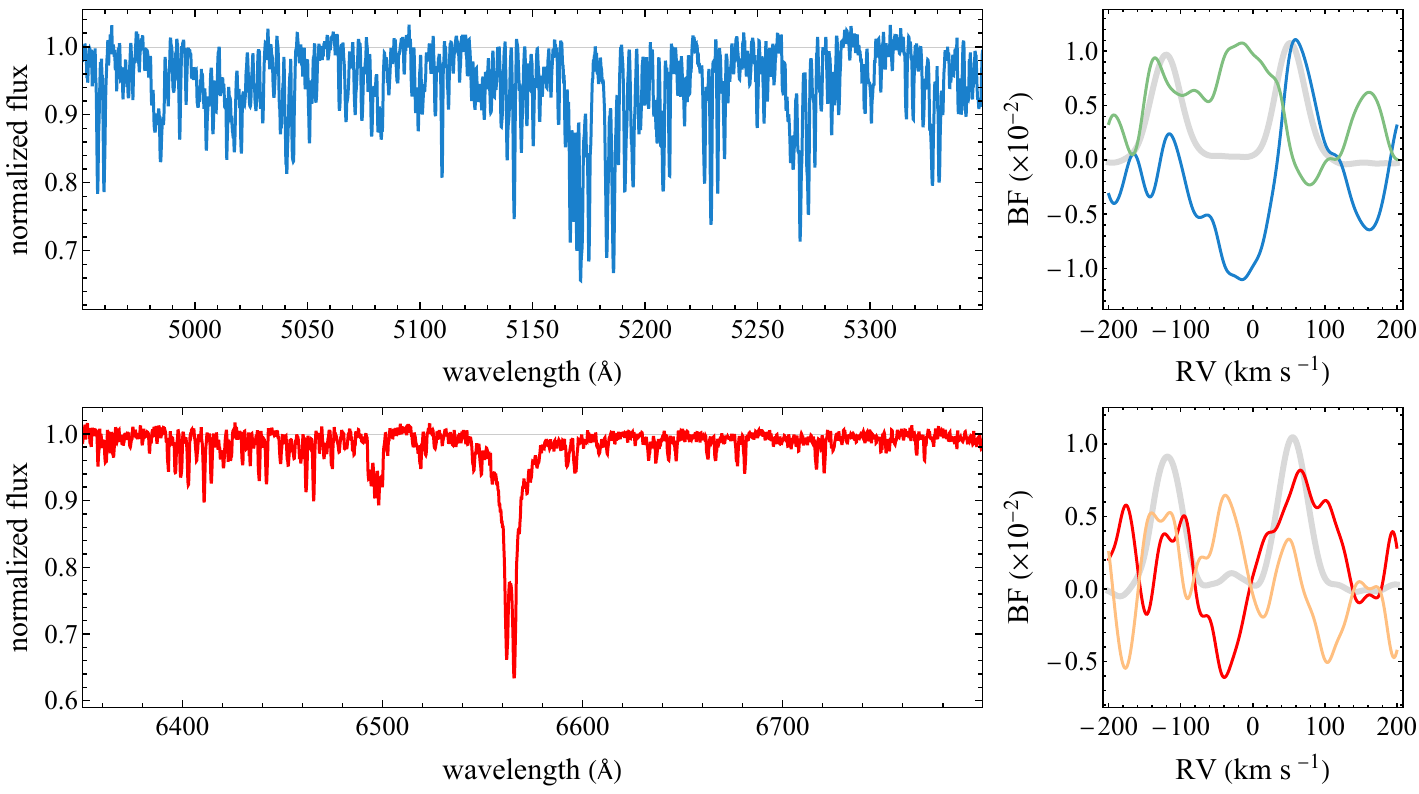}
    \caption{A failed case for the BBF. 
    Upper left panel: medium-resolution spectrum of LAMOST J085118.08+134541.0 on the blue side.
    Upper right panel: the BBF (blue and green lines) and the BF (shallow grey line) for the blue side spectrum.
    Lower left panel: medium-resolution spectrum of LAMOST J085118.08+134541.0 on the red side.
    Lower right panel: the BBF (red and orange lines) and the BF (shallow grey line) for the red side spectrum.
    In this case, the SB2 has almost identical components, the BBF fails to decompose the two components.
    Using the BF (shallow grey lines) as a visual guide, 
    one can find that the individual BFs contain non-physical wiggles that entangle with each other due to the degeneracy of the parameters.}
    \label{fig:fail}
\end{figure*}

\subsection{Issue about the Choice of Templates }\label{sec:temp}

In practice, one cannot find templates with perfectly matched stellar parameters to the target.
The variation of BBF to the choice of templates is explored in this section.
To this end, we use the mock SB2 described in Section~\ref{sec:sb2} and run a set of `bias tests' as follows.

We first run `bias tests' by varying both templates' $T_\mathrm{eff}$ 
(that is, to mimic an under/overestimated $T_\mathrm{eff}$ for a chosen template) 
while keeping $\log{g}$ and $\mathrm{[Fe/H]}$ fixed at perfectly matched values.
The $T_\mathrm{eff}$ for either component can go up (for overestimated case) or down (for underestimated case) by 100 K per check.
Figure~\ref{fig:teff} shows the variation of BBFs corresponding to five $\times$ five pairs of $\Delta T_\mathrm{eff}$
(each panel for one pair). 
For example, panel (23) in the second row and third column shows results for 
an underestimate of the primary's temperature by 100 K ($\Delta T_\mathrm{eff~1} = -100$ K)  
while a perfect match for the secondary's temperature ($\Delta T_\mathrm{eff~2} = 0$ K);
panel (41) in the fourth row and first column shows results for 
an overestimate of the primary's temperature by 100 K ($\Delta T_\mathrm{eff~1} = +100$ K)  
but an underestimate for the secondary's temperature by 200K ($\Delta T_\mathrm{eff~2} = -200$ K).
A clear pattern can be observed through the panels. 
Under/overestimating the $T_\mathrm{eff~2}$ induces a bump/pit on $\mathrm{BF_{1}}$ right below the peak of $\mathrm{BF_{2}}$ 
(by looking through the third row of the panels);
and under/overestimating the $T_\mathrm{eff~1}$ induces a bump/pit on $\mathrm{BF_{2}}$ right below the peak of $\mathrm{BF_{1}}$
(by looking through the third column of the panels).
Apparently, none of these distortions should exist as an individual BF has a clearly defined peak and baseline.
The reason for a bump is to compensate the underestimated equivalent line width given a biased $T_\mathrm{eff}$
and the reason for a pit is to eliminate overestimated  line width given a biased $T_\mathrm{eff}$. 
When both  $T_\mathrm{eff~1}$ and  $T_\mathrm{eff~2}$ are under- and/or over-estimated, 
either a bump or pit shows up depends on whether the total line width of the templates is over-/under-matched to the observation.
The results show that BBF can be quite sensitive to over/underestimate of $T_\mathrm{eff}$. 

Similar tests are implemented by varying the templates' $\mathrm{[Fe/H]}$ 
while keeping $T_\mathrm{eff}$ and $\log{g}$ fixed at their perfectly matched values,
and by varying the  templates' $\log{g}$ while keeping $T_\mathrm{eff}$ and  $\mathrm{[Fe/H]}$ fixed at their perfectly matched values.
For biased $\mathrm{[Fe/H]}$s (Figure~\ref{fig:meta}), the results show a similar pattern with that of the biased $T_\mathrm{eff}$ case.
Distortions (bump/pit) on BFs suggest that the results are also highly sensitive to the choice of $\mathrm{[Fe/H]}$. 
The reason holds the same, as templates with under- and/or over-estimated $\mathrm{[Fe/H]}$s and therefore line widths cannot otherwise fit the data well.
For biased $\log{g}$s however, we see barely distortions on components' BFs (Figure~\ref{fig:logg}).
The reasons are that: (1) the gravity-dependence of the line width are much weaker than the temperature dependence \citep[][ page 322]{Gray2005},
and (2) the mock spectral types we created are both cool stars ($< 10000$ K),
on which the line width have little gravity-dependence \citep[][ page 368]{Gray2005}.

The above tests suggest that the BBF may be used to constrain the stellar parameters for individual components of an SB2.
For instance, by examining the distortions on BFs, one can infer how well/bad the templates are chosen. 
Inspired by the work of \cite{Simon1994}, 
we calculate the residuum as \texttt{res} $ = ||\hat{\mathrm{D}} \vec{\mathrm{B}} -  \vec{\mathrm{S}}|| $ for each bias-test
and record the results in each corresponding panel 
(Figures~\ref{fig:teff}--\ref{fig:logg}; $\texttt{res}_{1}$ is the residuum for the BF and $\texttt{res}_{2}$ is the residuum for the BBF).
It is not surprising that the residuum is minimized when using templates with perfectly matched parameters
and the residuum increases when the parameters are increasingly biased. 
This hints that one can utilize optimization algorithms such as the Nelder-Mead algorithm to minimize the residuum 
and to find the best-match (least-square) templates.
We test this method again using the mock SB2 for the partially blended case (Figure~\ref{fig:bf123}; middle panel).
Mock observational spectra with SNR = 10, 20, 50, 100, 200, and 500 are simulated and tested. 
We set boundaries for each of these parameters as follows:
$T_\mathrm{eff~1} \in [5000, 6000]$ K,  
$\log{g}_{1} \in [3, 5]$ dex, 
$\mathrm{[Fe/H]}_{1} \in [-0.5, 0.5]$ dex,
$T_\mathrm{eff~2} \in [4000, 5000] $ K,  
$\log{g}_{2}  \in [1, 4]$ dex, and
$\mathrm{[Fe/H]} _{2} \in [-0.5, 0.5]$ dex.
In each iteration, the templates for the primary and secondary are interpolated from the BOSZ model grid (Section~\ref{sec:sb2})
given a set of stellar parameters at current iteration. 
The results show that all the six parameters can converge to values in agreement with the correct values 
used to generate mock SB2 (Figure~\ref{fig:optim}), even when the SNR is low. 
In comparison, we also run parallel tests using the single-template BF wrapped up with Nelder-Mead.
The best single-template's parameters tend to run close to that of the primary.
However, as mentioned, the result best-fit template does not represent secondary nor primary, 
For instance, the best-fit $T_\mathrm{eff} \simeq 5550$ K falls in between that of 
the primary ($T_\mathrm{eff~1} \simeq 5800$ K) and secondary ($T_\mathrm{eff~2} \simeq 4500$ K),
which represents but a flux-weighted template from the two components.

\subsection{Pros and Cons of the BBF}

The BBF inherits all the advantages of the BF
such as linearity, preservation of resolution, and well-defined peaks and baseline.
Taking one step further, the BBF can provide BFs for individual components without prior knowledge of underlying line profiles.
In principle, if one knows well the underlying line profiles model, 
he/she can simply fit the models to the BF derived using single template to get decomposed BFs.
However, in real cases the line profile models are typically not accessible due to:
blended broadenings from rotation and micro/macro-turbulent or instrumental broadening;
distorted stellar shapes (ellipsoidal effects commonly seen in close binaries), 
or subtle effects like cold/hot spots and eclipsing. 
Therefore we suggest to obtain individual BFs in a model independent way to first gain insights on the underlying physic, 
then proceed with careful modelling to the problem.

By including a second template, there are a few caveats introduced.
First, the number of columns and thus the size of the design matrix is doubled, 
making the BBF more computationally expensive than the BF.
Second, the number of unknowns are doubled, 
reducing the overdeterminacy by two-folds given a same input spectrum.
Now we have two BFs of a same length as the single BF case, 
therefore the SNR of individual BFs will decrease.
This issue becomes more of a problem if the SNR of a spectrum is poor.  
To restore SNR, one has two options.
The first option is to smooth the result BFs, for instance, by convolving a gaussian filter. 
In the case of low SNR, the FWHM of this gaussian filter would require a few times the spectral resolution,
thus the resolution and the shape of the BF will be sacrificed, 
sharp features (for instance, pits induced by spots) and broadening information may be lost.
The second option is to include more useful pixels of the input spectrum by enlarging the wavelength window.
However, too large of a wavelength window (for instance $\gtrsim$ 1000 \AA) 
will resulting a less localized, but somewhat averaged BF,
since the relative flux ratio of the two components varies across the wavelength window (see more discussion in the next paragraph).
In addition, for some instruments with limited wavelength coverage, this approach is simply not accessible. 
These caveats naturally becomes more severe if one tries to apply the BBF on triple or multiple systems.

\subsection{Precautions for Using the BBF}\label{subsec:caution}

One important aspect to understand and implement Equation~(\ref{eq:bf2}) should be noted.
Strictly speaking, one must use un-normalized (rather than normalized) templates 
to construct the design matrices $\hat{\mathrm{D}}_\mathrm{1, 2}$ (see Section~\ref{sec:bbf2}). 
The reason is that any two components with unequal spectral types
have different continuum slopes for a given finite wavelength window (that is, continua's flux ratio varies along the wavelength).
An actual SB2's spectrum is the superposition of un-normalized spectra of components.
If one uses normalized templates to model the problem, the variation of the continua's flux ratio is wiped out.
The good news is that since the flux ratio variation is typically not significant for a wavelength window spanning from $\sim$100 \AA\, up to 1000 \AA\,,
using normalized templates still works well as demonstrated in this work.
However, one must keep in mind that this issue becomes more of a problem 
when implementing the method on a larger wavelength window like $\gtrsim$ 1000\AA.
The recipe for adapting the variation is straight forward. 
In the first step, one constructs the design matrices $\hat{\mathrm{D}}_\mathrm{1, 2}$ using same procedures described in Section~\ref{sec:bbf2},
except that the columns of $\hat{\mathrm{D}}_\mathrm{1, 2}$ are now filled with un-nomalized flux vectors of given templates.
Therefore the slopes and hence the flux ratio of the continua are preserved. 
One can sum up the continua to obtain the total continuum level for the system, which will be used in the next step.
In the second step, one has to rectify/rescale/recalibrate the observation spectrum $\mathcal{S}$ onto the total continuum of the templates. 
In other words,  one has to fit typically uncalibrated continuum of $\mathcal{S}$ onto the sum of the two templates' continua.
The rectification can be achieved by first taking a pseudo-continuum normalisation for $\mathcal{S}$ 
then multiplying the normalized $\mathcal{S}$  by a lower-order best-fit polynomial to the total continuum obtained in the previous step.
Once $\mathcal{S}$ is properly rescaled, the BBF can then be solved using the singular value decomposition technique.
Since in this case, the flux ratio is determined once the templates are chosen,  
the derived component BBFs will be automatically normalized independently,
that is, the area under each component BF is summed up to unity.

The BBF works well for SB2s with a significant deviation of spectral types for the components.
Specifically, when the two components have a significant deviation of temperature,
the deviation of the line strength lifts the degeneracy and makes the decompose of individual BF feasible.   
For SB2s with similar or nearly identical components, 
one should not (there is no need to) include two templates for the problem,
as the BBF may fail to distinguish the two due to the degeneracy introduced by the (nearly) identical spectral lines.
As a demonstration, we tried to implement the BBF on an actual SB2 designated LAMOST J085118.08+134541.0.
The system was observed by a LAMOST -- K2 mission (refer to details in \cite{Wang2021})
with multi-epoch medium-resolution spectroscopy ($R\sim7500$) available from LAMOST DR10\footnote{http://www.lamost.org/dr10/}.
LAMOST J085118.08+134541.0 contains two similar stars with a slight difference on the effective temperature.
The parameters for the slightly hotter star are: 
$T_\mathrm{eff} \sim 6000$K,  $\log{g} \sim 4.1$ dex, $\mathrm{[Fe/H]} \sim -0.5$ dex;
and the parameters for the slightly cooler star are:
$T_\mathrm{eff} \sim 5800$K,  $\log{g} \sim 4.1$ dex, $\mathrm{[Fe/H]} \sim -0.5$ dex.
Figure~\ref{fig:fail} shows LAMOST J085118.08+134541.0's spectra and 
the decomposed BFs where the templates are set to the parameters given above.
The spectra show clear double-lined feature, with nearly equal line-depth. 
In this case the BBF fails to decompose the two even when the RV offset is large.
As a visual guide, we plot the result for the single-template BF for comparison.
It is no doubt that the BF should be used instead of the BBF, when the two components have similar parameters.
Empirically, we found that the BBF performs well for SB2s containing two components
with a large temperature difference (e.g.,  $\Delta T_\mathrm{eff} \gtrsim$ 1000 K );
or with similar temperatures, but very different luminosities and line strengths like some overcontact binaries,
although the specific successful rate of obtaining well-decomposed BBF could also depend on the SNR, spectral resolution, 
as well as the spectral types of the binary stars.

\section{Conclusions}\label{sec:conclude}

In this work, we proposed an extension to the BF by allowing the algorithm to take two/multiple templates, 
thus obtaining individual BFs for components in binary/multiple systems. 
We tested the BBF on mock SB2s and actual SB2s observed by APOGEE. 
Our results show that the method is feasible for SB2s even when the components are heavily blended. 
The potential of using the BBF to image transiting circumbinary exoplanets is studied and a few technical aspects of implementing the method are discussed. 
More applications of the BBF are expected in the future with high-resolution spectroscopy of binary/multi-star systems.


\begin{acknowledgments}
The author would like to thank the following: 
Subo Dong, Zexuan Wu, Huawei Zhang, Huiling Chen, Song Wang, and Rucinski S. M. for enlightening discussions,
Zhi-Xiang Zhang for providing expertise in stellar spectroscopy, 
Mouyuan Sun for providing instructions on mock spectroscopy experiments,
and the anonymous referee for providing beneficial suggestions that improved the paper.
This work is supported by the fellowship of China National Postdoctoral Program for Innovation Talents under grant BX20230020.

Funding for the Sloan Digital Sky Survey IV has been provided by the Alfred P. Sloan Foundation, the U.S. Department of Energy Office of Science, and the Participating Institutions. SDSS acknowledges support and resources from the Center for High-Performance Computing at the University of Utah. The SDSS website is www.sdss.org.

SDSS is managed by the Astrophysical Research Consortium for the Participating Institutions of the SDSS Collaboration including the Brazilian Participation Group, the Carnegie Institution for Science, Carnegie Mellon University, the Chilean Participation Group, the French Participation Group, Harvard-Smithsonian Center for Astrophysics, Instituto de Astrofísica de Canarias, The Johns Hopkins University, Kavli Institute for the Physics and Mathematics of the Universe (IPMU)/University of Tokyo, Lawrence Berkeley National Laboratory, Leibniz Institut für Astrophysik Potsdam (AIP), Max-Planck-Institut für Astronomie (MPIA Heidelberg), Max-Planck-Institut für Astrophysik (MPA Garching), Max-Planck-Institut für Extraterrestrische Physik (MPE), National Astronomical Observatories of China, New Mexico State University, New York University, University of Notre Dame, Observatório Nacional/MCTI, The Ohio State University, Pennsylvania State University, Shanghai Astronomical Observatory, United Kingdom Participation Group, Universidad Nacional Autónoma de México, University of Arizona, University of Colorado Boulder, University of Oxford, University of Portsmouth, University of Utah, University of Virginia, University of Washington, University of Wisconsin, Vanderbilt University, and Yale University.

Guoshoujing Telescope (LAMOST) is a National Major Scientific Project built by the Chinese Academy of Sciences. 
Funding for the project has been provided by the National Development and Reform Commission. 
LAMOST is operated and managed by the National Astronomical Observatories, Chinese Academy of Sciences.
\end{acknowledgments}

\clearpage

\bibliography{reference}{}
\bibliographystyle{aasjournal}


\end{CJK*}
\end{document}